# Chloride Molecular Doping Technique on 2D Materials: $WS_2$ and $MoS_2$

Lingming Yang[1], Kausik Majumdar[2], Han Liu[1], Yuchen Du[1], Heng Wu[1], Michael Hatzistergos[3], P. Y. Hung[2], Robert Tieckelmann[2], Wilman Tsai[4], Chris Hobbs[2], Peide D. Ye[1*]

[1]School of Electrical and Computer Engineering and Birck Nanotechnology Center, Purdue University, West Lafayette, IN 47907, U.S.A.
[2]SEMATECH, Albany, NY 12203, U.S.A.
[3]College of Nanoscale Science and Engineering, State University of New York at Albany, Albany, NY 12203, U.S.A.
[4]Intel Corporation, Santa Clara, CA 95054, U.S.A.

**\*Corresponding Author**
Prof. Peide D. Ye
School of Electrical and Computer Engineering and Birck Nanotechnology Center, Purdue University, West Lafayette, IN 47907, USA
Tel: 1-765-494-7611, Fax: 1-765-496-7443
E-mail: yep@purdue.edu

**Abstract:**

Low-resistivity metal-semiconductor (M-S) contact is one of the urgent challenges in the research of 2D transition metal dichalcogenides (TMDs). Here, we report a chloride molecular doping technique which greatly reduces the contact resistance ($R_c$) in the few-layer $WS_2$ and $MoS_2$. After doping, the $R_c$ of $WS_2$ and $MoS_2$ have been decreased to 0.7 k$\Omega\cdot$μm and 0.5 k$\Omega\cdot$μm, respectively. The significant reduction of the $R_c$ is attributed to the achieved high electron doping density thus significant reduction of Schottky barrier width. As a proof-of-concept, high-performance few-layer $WS_2$ field-effect transistors (FETs) are demonstrated, exhibiting a high drain current of 380 μA/μm, an on/off ratio of 4×10$^6$, and a peak field-effect mobility of 60 cm$^2$/V$\cdot$s. This doping technique provides a highly viable route to diminish the $R_c$ in TMDs, paving the way for high-performance 2D nano-electronic devices.

**Introduction:**

Recently 2D TMD materials have trigged intensive research interests due to their unique electrical,[1,2] optical,[3] and mechanical properties.[4] $MoS_2$, one of the most studied TMD materials, has been used as the channel material of the FET, exhibiting high on/off ratio and high mobility.[5-9] Recent studies on the transistor characteristics of $MoS_2$ FETs reveal that the $MoS_2$ transistors are essentially Schottky barrier transistors whose switching is controlled by the tuning of the Schottky barrier at the contacts[10]. The model of Schottky barrier transistor also holds true in other TMD materials such as $MoSe_2$,[11] $WS_2$,[12] and $WSe_2$.[13] For Schottky barrier transistors, the intrinsic properties of the TMDs channel are masked by the Schottky contacts and no merits are gained from aggressive scaling.[14] In order to deal with the Schottky contact issue in TMDs, intensive research efforts have been made in recent years. A lot of attention has been paid to various contact

metals including Sc,[15] In,[13] Al,[13, 16] Ti,[10, 13, 15, 17] Cr,[8] Mo,[18] Ni,[11, 15, 19], Au,[7, 16] and Pt.[15, 16] Unfortunately, most of these M-S contacts showed non-negligible Schottky barriers regardless of the contact metals.[16] Among them, some studies showed interesting results by engineering the contact metal. For instance, the Sc-MoS$_2$ contact shows a very low M-S interface resistance (0.65 kΩ·µm). However, the overall R$_c$ is still considered to be large (~5 kΩ·µm) due to the semiconductor internal resistance.[15] Other studies were aimed to metalize the TMDs under the contact by forming chemical bonds between sulfur and contact metal. W. Liu *et al.* reported an R$_c$ of 0.8 kΩ·µm in Ti-15-layers-MoS$_2$ but the R$_c$ of Ti-1-layer-MoS$_2$ is as large as 740 kΩ·µm.[17] In general, it is difficult to achieve a low R$_c$ in TMDs by simply using contact metals with low work functions since the Fermi-level tends to be pinned at charge neutrality level (CNL) or S-vacancy level which is located below the conduction band edge with non-negligible Schottky barrier heights.[20, 21] To make things worse, it's even more difficult to achieve low R$_c$ in other TMD materials (such as WS$_2$) because their CNLs are located in the middle of the bandgap[22, 23] with larger Schottky barriers for electrons compared with MoS$_2$.[12]

Another key way to achieve the low R$_c$ on the M-S contact is to heavily dope the semiconductor under the metal.[24] After heavily doping, the Schottky barrier width is reduced and current through the M-S contact is greatly enhanced by electron tunneling. The doping process is achieved by dopant diffusion or ion-implantation in traditional semiconductors. For the atomically thin semiconductors, it is a challenge to precisely control the doping density with the ion-implantation since the thickness of the semiconductor is only a few nanometers.[25] On the other hand, new doping techniques, such as molecular doping, show the potential advantages over ion-implantation when applied to the ultrathin 2D semiconductors. H. Fang *et al.* used potassium[26] as an adatom dopant to reduce the R$_c$ in WSe$_2$ and MoS$_2$. Y. C. Du *et al.* also explored to apply PEI molecular doping to reduce the $R_c$ in MoS$_2$.[27] However, the aforementioned two methods have limited improvement in the $R_c$ and also doping effect degrades with time. In this letter, we report a novel chloride molecular doping method on TMD materials. After doping, the $R_c$ of Ni-WS$_2$ and Ni-MoS$_2$ contacts has been significantly reduced to 0.7 kΩ·μm and 0.5 kΩ·μm, respectively. High doping density of $6.0 \times 10^{11}$ cm$^{-2}$ and $9.2 \times 10^{12}$ cm$^{-2}$ are achieved on the few-layer WS$_2$ and MoS$_2$ at zero back gate biases. XPS results reveal that the high doping densities are attributed to the doping by Cl. In addition, high performance few-layer WS$_2$ FETs have been demonstrated with a drain current of 380 μA/μm, an $I_{on}/I_{off}$ ratio $> 4\times10^6$, and a peak field-effect mobility of 60 cm$^{-2}$/V·s. No degradation of $I_{on}/I_{off}$ ratio is observed after the doping. The results show that the chloride molecular doping method is a promising way to achieve low $R_c$ in TMD materials.

**Results and Discussion:**

Figure 1a schematically shows the structure of Cl-doped few-layer WS$_2$ or MoS$_2$ back gate FETs. The fabrication process starts with the mechanical exfoliation of bulk WS$_2$ (Nanosurf Inc.) and bulk MoS$_2$ (SPI supplies) using scotch tape method. The exfoliated WS$_2$ or MoS$_2$ flakes are transferred onto a 90 nm SiO$_2$/p$^{++}$ Si wafer and then soaked in undiluted 1, 2 dichloroethane (DCE) (99.8%, SIGMA-ALDRICH, No. 34872) at room temperature for more than 12 hours. The thickness is about 3.5 to 5 nm (5-7 monolayer) which is first identified by optical microscope and measured by Atomic Force Microscopy (AFM). Using thin flake (1-4 layers) will lead to higher contact resistance due to the increasing of the bandgap (see the Supporting Information). Acetone and isopropanol rinses are performed to remove the residual chemical. E-beam lithography is used to define the contact region with a fixed contact width of 1 μm. Transfer length method (TLM) structures with various gap spaces are designed to extract the $R_c$. The combination of Ni (30 nm)/Au (60 nm) is deposited as contact metals by e-beam evaporator at the pressure of $2\times10^{-6}$

Torr. The electrical measurements are carried out with Keithley 4200 Semiconductor Parameter Analyzer.

In order to verify the n-type doping, X-ray photoelectron spectroscopy (XPS) surface analysis is used to measure the Fermi levels in TMDs before and after the DCE treatment. As shown in Figure 1b, blue shifts (increase in energy) are observed in the binding energy of core levels in $WS_2$ after the DCE treatment. Because the binding energy of XPS spectra is referenced to Fermi level in the material, the blue shifts of binding energy can be interpreted by the move up of the Fermi level in the semiconductor. Figure 1c summarizes the energy shifts of core levels in $WS_2$ and $MoS_2$. For $WS_2$, the binding energy of $W4f_7$ shifts from 32.76 eV to 33.09 eV while the peak of $S2p_3$ shifts from 162.49 eV to 162.79 eV; for $MoS_2$, the binding energy of $Mo3d_5$ shifts from 228.66 eV to 229.42 eV while $S2p_3$ shifts from 161.39eV to 162.22 eV. An increment of about 0.3 eV and 0.8 eV in binding energy was observed in DCE treated $WS_2$ and $MoS_2$, respectively. It should be noticed that a larger shift of the binding energy in $MoS_2$ does not necessarily correspond to a larger shift of the Fermi level in it as the time lag between DCE treatment and XPS is different for two cases. The stability of the Cl-doped $MoS_2$ and $WS_2$ FETs can be seen in the Supporting Information. On the other hand, the threshold voltage of $WS_2$ and $MoS_2$ FETs exhibit a negative shift after DCE treatment. The negative shift of threshold voltage indicates there is a higher electron density in the semiconductor channels with DCE treatment. As a result, it is reasonable to conclude that the n-type doping is achieved by the DCE treatment. The doping mechanism on TMDs by the DCE treatment is not completely established yet. One possible mechanism is that the substitute doping of the TMDs is realized through the replacement of S vacancy by Cl atom. It has been reported that none of the element of halogen family is an effective dopant when acts as an adatom dopant.[28] However, previous simulation shows that when high density sulfur atoms are substituted by Cl atoms the discrete impurity energy levels in MoS$_2$ broaden into a band and merge with conduction band, resulting the band gap narrowing and the degeneration doping.[29] On the other hand, it was also observed that a significant amount of Cl element was detected on the flakes' surface by Secondary Ion Mass Spectrometry (SIMS) even though the TMD materials are thoroughly rinsed by acetone and isopropanol.[30] Given that sulfur vacancies are widely detected from the mineral MoS$_2$,[21] it is reasonably assuming that the doping effect is achieve by the extra electrons donated by Cl atoms when they occupy the location of sulfur vacancies.

The R$_c$ of WS$_2$ and MoS$_2$ can be significantly reduced after the Cl doping. Known as an ambipolar semiconductor, the undoped WS$_2$ shows large Schottky barriers for both electrons and holes, resulting an extremely large R$_c$.[12] For such a larger Schottky barrier, it would be impractical to extract the R$_c$ by the TLM structure which is applicable to Ohmic or low resistivity contacts only. However, a simple estimation of the R$_c$ of the undoped WS$_2$ is on the order of $10^2$ kΩ·μm since the total resistance of the 100 nm device is calculated to be $5\times10^2$ kΩ·μm. After doping, an R$_c$ as low as 0.7 kΩ·μm, 2-3 orders of magnitude reduction, can be extracted by linearly fitting the curve of total resistances. Figure 2a shows the TLM resistances of the Cl-doped WS$_2$ and MoS$_2$ as a function of gap space at a back gate bias of 50 V. The corresponding transfer length ($L_t$) of Cl-doped WS$_2$ is about 132 nm extracted from the TLM. The specific contact resistivity ($\rho = R_c L_T W$) is calculated to be $9.2 \times 10^{-6}$ Ω·cm$^2$, where $W$ is the channel width. The sheet doping density in WS$_2$ is about $6.0 \times 10^{11}$ cm$^{-2}$ determined by the equation of $n = \frac{L}{eWR_{sh}\mu}$, where $L$ is the channel length, $e$ is the electron charge, $R_{sh}$ is the sheet resistance, and $\mu$ is the field-effect mobility shown in Figure 4a. To the best of our knowledge, such a low R$_c$ has never been achieved on WS$_2$ or other TMDs whose CNL is located in the middle of the bandgap. On the other hand, an even lower R$_c$ of 0.5 kΩ·μm was obtained on Cl doped MoS$_2$. Compared with the R$_c$ of the undoped

MoS$_2$ (about 5-6 kΩ·μm),[19] a ten times of reduction has also been achieved after doping.[30] The $\rho_c$ and the doping density of Cl-doped MoS$_2$ are determined to be 3×10$^{-7}$ Ω·cm$^2$ and 9.2 × 10$^{12}$ cm$^{-2}$, respectively.

The mechanism of the reduction of the R$_c$ on Cl-doped TMDs is very clear. The energy band diagrams of the Ni-WS$_2$ and Ni-MoS$_2$ contacts with and without the Cl doping are shown in Figure 2b. The Fermi level at the metal-WS$_2$ interface is pinned near the CNL, resulting a significantly large Schottky barrier. The barrier height is given by $\Phi = \lambda_{gs} (\Phi_m - \chi_s) + \Phi_0 (1 - \lambda_{gs})$,[31] where $\lambda_{gs}$ is gap states parameter, $\Phi_m$ is the metal work function, $\chi_s$ is the electron affinity of the semiconductor, and $\Phi_0$ is the charge neutrality position. The height of the Schottky is large enough to rectify the electrons' ejection from the metal to the semiconductor at low $V_{ds}$, as shown in Figure 2b. Moreover, this barrier height can't be efficiently modified by varying the workfunction of contact metals due to the complicated metal-to-TMD interface.[21] The difference of the R$_c$ between WS$_2$ and MoS$_2$ is due to the difference alignment of the CNL in the two materials.[22, 32] Compared with MoS$_2$, the CNL in WS$_2$ is more close to the mid of bandgap, resulting a larger Schottky barrier. Without doping, it would be much harder for the electrons to inject from the metal to the semiconductor in WS$_2$ because the thermionic current exponentially decreases with the increasing of barrier height. However, when the tunneling current starts to dominate the current through the M-S junction, the electron injection through the barrier becomes much easier. The effective electron density (induced by chemical doping and electrostatic doping) at $V_{bg}$ of 50 V is as high as 2.3 × 10$^{13}$ cm$^{-2}$ and 2.9 × 10$^{13}$ cm$^{-2}$ for WS$_2$ and MoS$_2$, respectively. As a result, both of the R$_c$ in the WS$_2$ and MoS$_2$ decrease significantly after doping. However, it is interesting that most of the electron density in WS$_2$ is attributed to the back gate bias rather the chemical doping. In other word, the Fermi level (electron density) at the interface can be effectively modulated by the back gate bias. Effective modulation via field-effect can be ascribed to the passivation of sulfur vacancy by Cl, given that the sulfur vacancy is the cause of the Fermi level pining on $MoS_2$ and $WS_2$ at M-S interface.[20] The hysteresis of the I-V characteristics due to the doping ions has also been discussed in the Supporting Information.

It is reasonable to predict that the molecular doping method would also be valid in other TMDs as long as there are vacancies of chalcogenide elements. The reduction of the $R_c$ by the doping technique is extremely important for TMDs whose Fermi level is pinned close to the mid bandgap. Due to the relative low Schottky barrier, $MoS_2$ has been given overwhelming attentions in recent years. However, for most of the TMDs, their intrinsic properties have not be fully investigated because they are hidden by the large $R_c$.[10] By effectively reducing the $R_c$, the molecular doping technique can be applied to other TMDs research. Applying low work function metal Ti is not successful. We ascribe the failure to the fact that Ti is a getter which deteriorates the substitute doping of Cl.

Since the low $R_c$ is achieved in $WS_2$ by Cl doping, high-performance $WS_2$ FET is expected. The output characteristics of the Cl-doped few-layer $WS_2$ FETs with 100 nm channel length are shown in Figure 3a. The device exhibits promising device performance including a drain current of 380 μA/μm as well as good current saturation. Due to a small $R_c$, the linear region of the $I_{ds}$-$V_{ds}$ curves shows excellent linearity. The drain current starts to saturate at $V_{ds}$ of 1.0 V due to the electron velocity saturation. Figure 3b shows the $I_{ds}$-$V_{ds}$ curves of the 100 nm $WS_2$ FET without Cl doping, which is used as a control device. Obvious rectify characteristics are observed at the low drain bias (< 0.5 V), indicating a large Schottky barrier at the contact. Compared with undoped devices, the drive current of the Cl-doped $WS_2$ FET has been improved by more than 6 times. Figure 3c shows the transfer curves of the Cl-doped few-layer $WS_2$ FET with 100 nm channel length. Due to a relative large band gap and ultra-thin-body-on-insulator structure, the off current is as low as $10^{-10}$ μA/μm and $10^{-12}$ μA/μm at drain biases of 2 V and 0.05 V, respectively. The low off-state current is attractive for low power applications especially when the device works at the ultra-scaled gate length. The $I_{on}/I_{off}$ ratio is about $4\times10^6$ and $3\times10^7$ at the drain bias of 2 V and 0.05 V, respectively. Compared with the undoped WS$_2$ FET in Figure 3d, the $I_{on}/I_{off}$ ratio has been increased by 2 folds due to the improvement of the on-current. By the linearly extrapolating of $I_{ds}$-$V_{gs}$ curves, the threshold voltage is calculated to be −14 V and 1.7 V for the doped and undoped devices, respectively. The negative shift of the threshold voltage is due to the n-type doping in channel. Unlike other doping methods which use charge transfer from absorbed molecules or atoms such as PEI and potassium, the off current of the Cl-doped FETs don't exhibit any degradation even at 100 nm short channel length. This result also indicates that the doping procedure is not fulfilled by the surface adhesion of extra atoms or molecules otherwise there would be a leakage current through the doping layer. The good device performance of Cl-doped WS$_2$ FETs shows that the presented chloride molecular doping techniques is a powerful method to be applied to dope the TMDs and fabricate high performance 2D FETs.

We have also investigated the scaling down trends of the Cl-doped WS$_2$ FETs, including the field-effect mobility, $I_{on}/I_{off}$ ratio and the drive current. Previously, the field-effect mobility of WS$_2$ was significantly underestimated due to the large Schottky barrier.[12, 31] As shown in Figure 4a, the field-effect mobility is carefully calculated by subtracting the R$_c$ at different back gate bias. The mobility is given by $\mu = (\frac{L}{WC_{ox}})\frac{d}{d(V_{bg}-V_{th})}(\frac{1}{R_{tot}-2R_c})$, where $W$ and $L$ are the channel width and length, $C_{ox}$ is the oxide capacitance, $V_{bg}$ is the back gate bias, $V_{th}$ is the threshold voltage, $R_{tot}$ is the total resistance, $2R_c$ is the total contact resistance. The peak field-effect mobility is about 60 cm$^2$/V·s, which is similar to the values in the Cl-doped MoS$_2$ FETs.[30] Considering the high doping density in channel, the motilities are among the good values in TMD materials. When the device is at high electron density region, the electron mobility decreases to around 20 cm$^{-2}$/V·s. At both the high $V_{bg}$ and low $V_{bg}$ regions, the obtained field-effect motilities agree well with the reported Hall mobility measured with ion liquid gate.[33] The scaling down trends of the drain currents of 0.1, 0.2, 0.5, 1 μm devices are shown in Figure 4b. The on current increases from 136 μA/μm to 380 μA/μm with the channel length scales down from 1 μm to 0.1 μm. Figure 4b also shows the $I_{on}/I_{off}$ ratio for devices with different channel lengths. For long channel devices, the $I_{on}/I_{off}$ ratios are more than $3 \times 10^7$. The $I_{on}/I_{off}$ ratio increases with the channel length scaling due to the increasing of the maximum drain current. For 100 nm devices, the off current increases because of the Drain-Induced-Barrier-Lowering. As the result, the $I_{on}/I_{off}$ ratio decreases to $4 \times 10^6$. Considering the thick gate oxide (90 nm SiO$_2$), it is reasonable to observe the degradation of the gate electrostatic control of the channel at the short gate length. If a thinner and high-k gate dielectric were used, a higher on/off ratio can be achieved. Previously, Cl-doped few-layer MoS$_2$ FETs have been demonstrated to have a low $R_c$ of 0.5 kΩ·μm, a high drain current of 460 μA/μm, and a high on/off ratio of $6.3 \times 10^5$.[30] Compared with MoS$_2$, Cl-doped WS$_2$ FETs exhibit comparable low $R_c$, high drain current, and field-effect mobility. The advantages of WS$_2$ FETs over MoS$_2$ ones are the low off-state current and the high on/off ratio.

**Conclusion:**

In summary, a novel chloride molecular doping technique is demonstrated on WS$_2$ and MoS$_2$. The $R_c$ of Ni-WS$_2$ and Ni-MoS$_2$ contacts have been reduced to 0.7 kΩ·μm and 0.5 kΩ·μm with the presented technique. The significant improvement in $R_c$ is due to the n-type doping thus the reduction of the Schottky barrier width. The doping mechanism could be the occupation of sulfur vacancies by chlorine atoms rather than the surface adhesion of chloride molecules. High performance few-layer $WS_2$ FETs have been successfully demonstrated using Cl doping. The 100 nm $WS_2$ FET has a high drain current of 380 μA/μm, a high on/off ratio of $4\times10^6$, and a field-effect mobility around 60 $cm^2/V·s$. In addition, no degeneration of $I_{on}/I_{off}$ ratio is observed after the Cl doping. The scaling down characteristics of the Cl-doped $WS_2$ FETs have also been investigated. Compared with Cl-doped $MoS_2$ FETs, Cl-doped $WS_2$ FETs show comparable electrical performance and better on/off ratio. This doping technique can also be widely applied to other TMD materials to achieve low $R_c$ and provide a route to realize high-performance electronic devices with 2D materials.

**ASSOCIATED CONTENT:**

Supporting Information Available: thickness dependence of the Cl-doped $MoS_2/WS_2$ FETs, hysteresis data, and stability of the device in vacuum and air. This material is available free of charge via the Internet at http://pubs.acs.org.

**ACKNOWLEDGMENT:**

The work at Purdue University is supported by SEMATECH and SRC. The authors would like to thank Hong Zhou, Yexin Deng and Zhe Luo for the valuable discussions and technical assistance.

**Figures:**

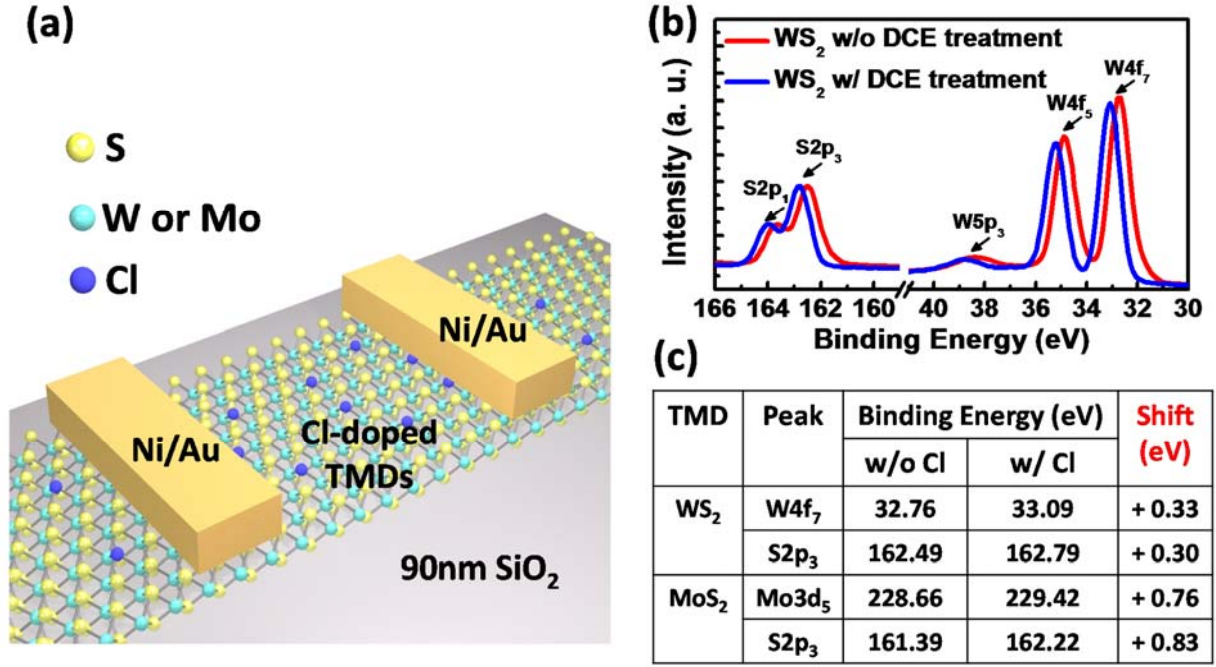

**Figure 1.** (a) Schematic of Cl-doped few-layer WS$_2$ back-gate FET. The chemical doping is achieved by soaking the flakes in the 1, 2 dichloroethane (DCE) solution for more than 12 hours and rinsed by acetone and isopropanol for 30 minutes. The contact metal Ni (30 nm)/Au (60 nm) is deposited immediately after the e-beam lithography. Back gate oxide is 90 nm SiO$_2$ and p$^{++}$ Si is used as back gate contact. (b) The binding energies of core levels in WS$_2$ with and without the DCE treatment. Blue shifts of the peak were observed after DCE treatment. (c) Summarize of the shifts of the binding energy in WS$_2$ and MoS$_2$ before and after DCE treatment. The energy peaks shows about + 0.3 eV shifts and + 0.8 eV shifts in WS$_2$ and MoS$_2$, respectively.

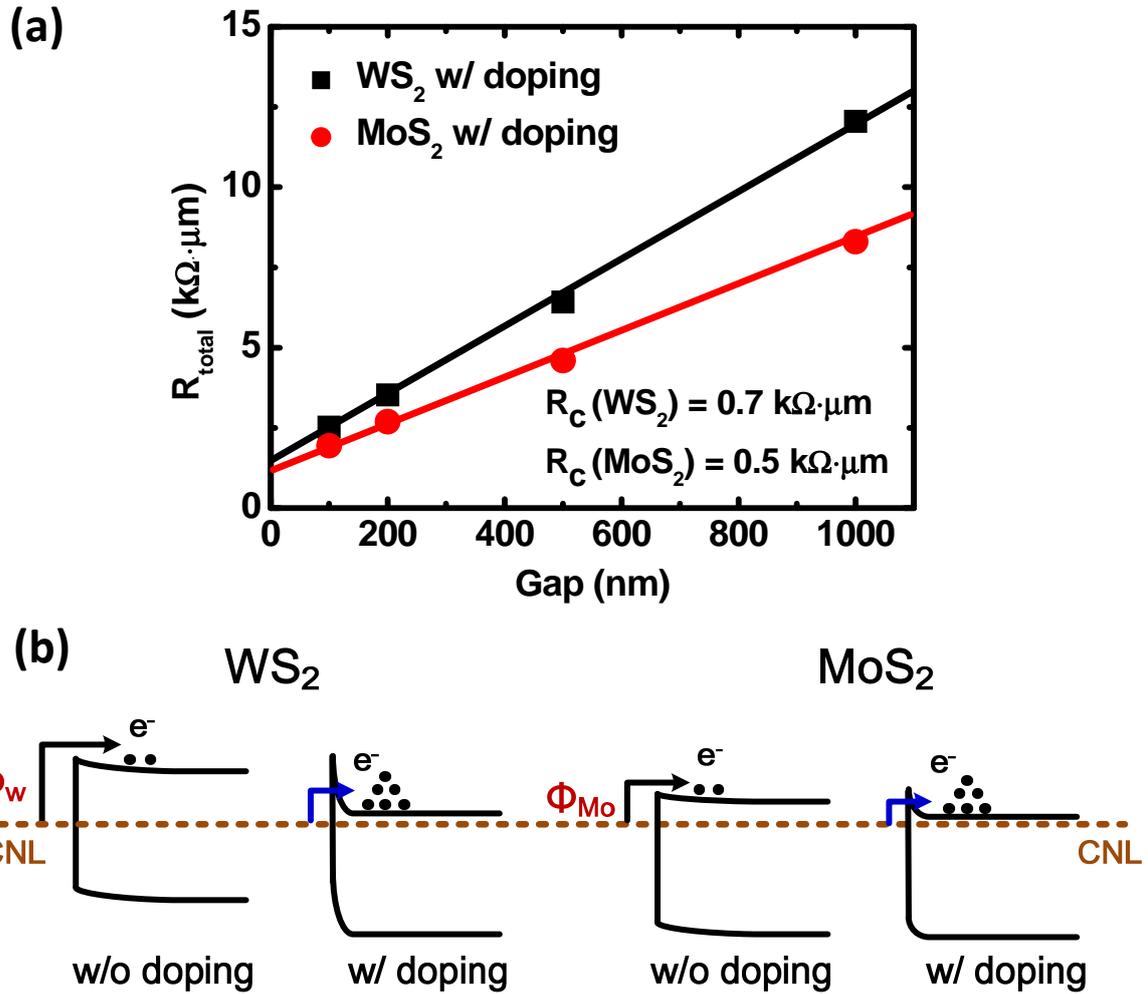

**Figure 2.** (a) TLM resistances of Cl-doped WS$_2$ and MoS$_2$. The R$_c$ is extracted to be 0.7 kΩ·μm and 0.5 kΩ·μm at the back gate bias of 50 V for WS$_2$ and MoS$_2$, respectively. For WS$_2$, the $L_T$ is extracted to be 132 nm and the corresponding ρ$_c$ is about 9.2×10$^{-6}$ Ω·cm$^2$ and the doping density is about 6.0 × 10$^{11}$ cm$^{-2}$. For MoS$_2$, the $L_T$ is 60 nm and the ρ$_c$ is about 3×10$^{-7}$ Ω·cm$^2$ and the doping density is about 9.2 × 10$^{12}$ cm$^{-2}$. (b) Schematic band diagram of metal-TMD contacts with and without chloride doping. Before DCE treatment, the Fermi level is pinned close to the CNL, resulting a large Schottky barrier. After DCE treatment, WS$_2$ and MoS$_2$ are heavily doped and the Fermi level in 2D materials can be efficiently moved after the passivation of S vacancy by Cl dopants.

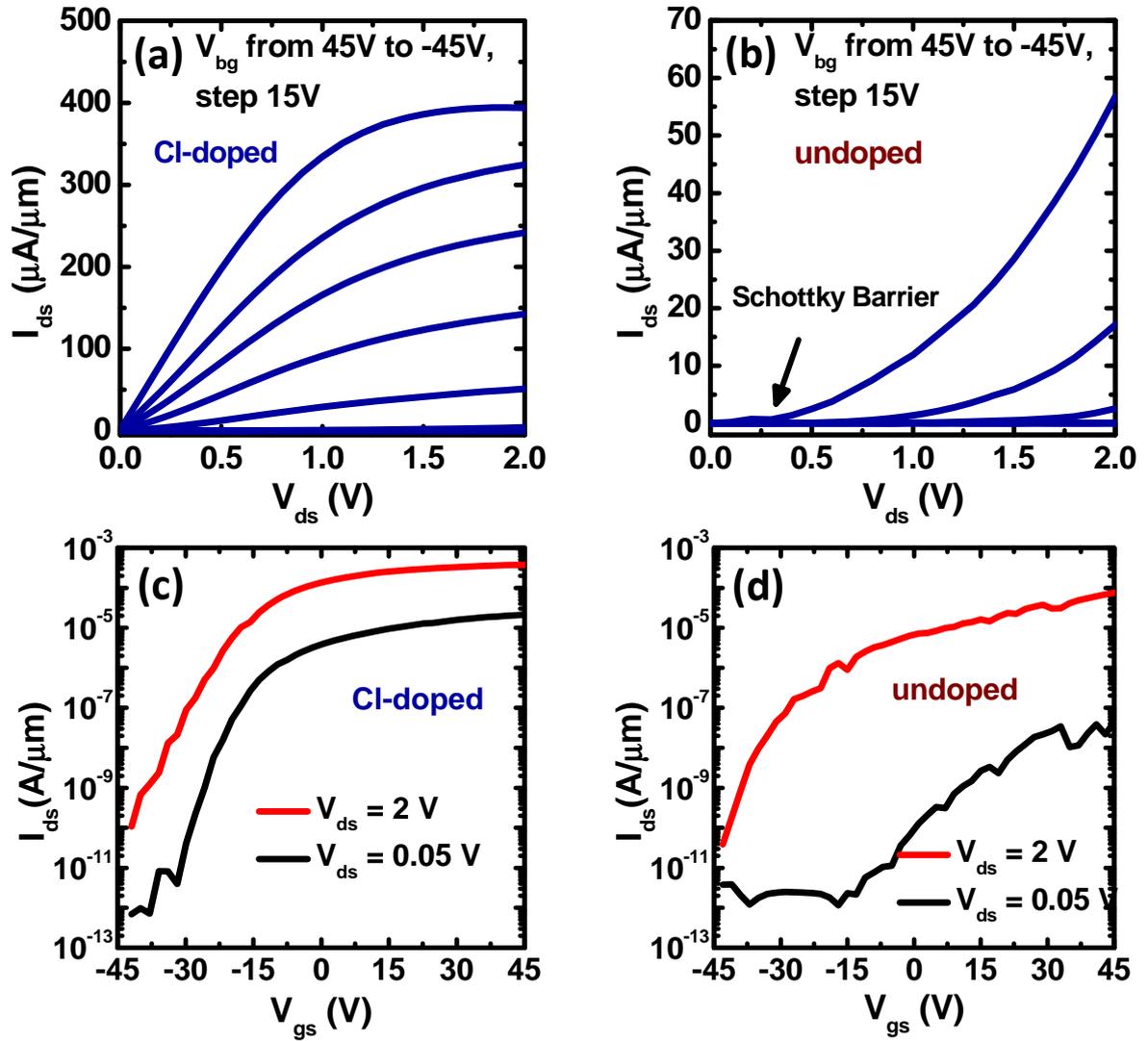

**Figure 3.** (a) Output characteristics of the Cl-doped few-layer WS$_2$ FETs at 100 nm channel length. The maximum drain current is enhanced to 380 µA/µm after Cl doping. Excellent current saturation is also observed. (b) Output characteristics of the undoped few-layer WS$_2$ FETs. The drain current is significantly suppressed by the Schottky barrier, indicating a larger barrier height. (c) Transfer characteristics of the device in Figure 3(a). The off current is as low as $10^{-10}$ µA/µm and $10^{-12}$ µA/µm at $V_{ds}$ of 2 V and 0.05 V. The I$_{on}$/I$_{off}$ ratio is about $4\times10^6$ and $3\times10^7$ at $V_{ds}$ of 2 V and 0.05 V, respectively. (d) Transfer characteristics of the device in Figure 3(b). The I$_{on}$/I$_{off}$ ratio is about $2\times10^6$ and $1.1\times10^4$ at $V_{ds}$ of 2 V and 0.05V.

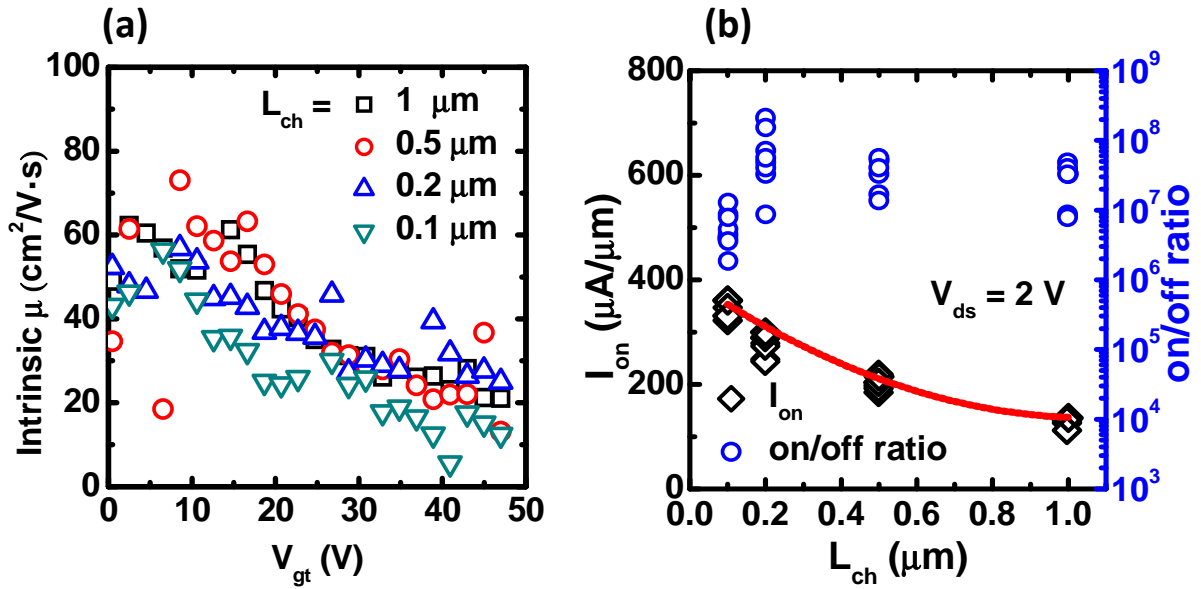

**Figure 4.** (a) The field-effect motilities as a function of the back gate bias for the WS$_2$ FETs with various channel lengths. The peak mobility is about 60 cm$^2$/V·s, which is close to the Hall mobility. (b) I$_{on}$/I$_{off}$ ratio and maximum drain current of Cl-doped few-layer WS$_2$ FETs as a function of channel length. For channel length > 0.2 μm, the I$_{on}$/I$_{off}$ ratio is more than 1×10$^7$; for 100 nm channel length device, the I$_{on}$/I$_{off}$ ratio decreases due to the loss of gate control. The drive current increases from 136 μA/μm to 380 μA/μm with the channel length scales down from 1 μm to 0.1 μm.

# Table of Content Graphic

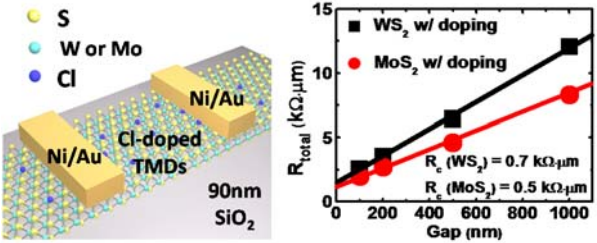

# Supporting Information for "Chloride Molecular Doping Technique on 2D Materials: WS$_2$ and MoS$_2$"

Lingming Yang[1], Kausik Majumdar[2], Han Liu[1], Yuchen Du[1], Heng Wu[1], Michael Hatzistergos[3], P. Y. Hung[2], Robert Tieckelmann[2], Wilman Tsai[4], Chris Hobbs[2], Peide D. Ye[1*]

[1]School of Electrical and Computer Engineering and Birck Nanotechnology Center, Purdue University, West Lafayette, IN 47907, U.S.A.
[2]SEMATECH, Albany, NY 12203, U.S.A.
[3]College of Nanoscale Science and Engineering, State University of New York at Albany, Albany, NY 12203, U.S.A.
[4]Intel Corporation, Santa Clara, CA 95054, U.S.A.

*Corresponding Author
Prof. Peide D. Ye
School of Electrical and Computer Engineering and Birck Nanotechnology Center, Purdue University, West Lafayette, IN 47907, USA
Tel: 1-765-494-7611, Fax: 1-765-496-7443
E-mail: yep@purdue.edu

**Thickness dependence of the Cl-doped FETs**

The thickness of the flakes used in this work is between about 3.5 to 5 nm (5-7 layers). The reason why we choose such a range of thickness is to balance the on and off currents so that the device performance is optimized. From the device point of view, the flakes with thinner thickness (1-4 layers) suffer from three disadvantage: 1, the larger band gap makes the Schottky barrier even larger, resulting a worse contact; 2, less charge screen lead to a degraded mobility compared with bulk mobility; 3, the charge density of the channel decreases and it limits the saturation drain current. As a result, both the channel resistance and the contact resistance become larger when a thinner flake is used as the channel. The main disadvantage of the thick flakes (>8 layers) is the difficulty to deplete the heavily doped layer on the top of the flake (which is away from bottom gate dielectric). As a result, the off current becomes larger if a thick flake is used. The layer-dependence of the characteristics of the $MoS_2$ and $WS_2$ FETs has also been experimentally investigated here. Figure S1 summarizes the drain current of the $MoS_2$ FETs versus the layer thickness. The $I_{ds}$-$V_{ds}$ curves of 100 nm channel-length Cl-doped $MoS_2$ and $WS_2$ FETs with 2-7 layers are shown in Figure S2 and Figure S4, respectively. The flake thicknesses were measured by AFM, as shown in Figure S3 and Figure S5. Compared with Cl-doped $MoS_2$, the performance of Cl-doped $WS_2$ FETs are significantly worse when the flake is less than 3 layers, which is consistent with the recently report on bilayer and monolayer $WS_2$.[1] The difference stems from the larger bandgap[2] and the larger Schottky barrier in $WS_2$.

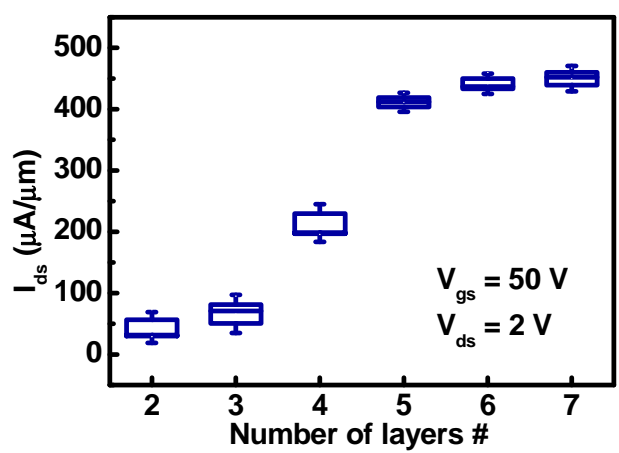

**Figure S1.** Drain currents of the Cl-doped $MoS_2$ FETs versus the number of layers.

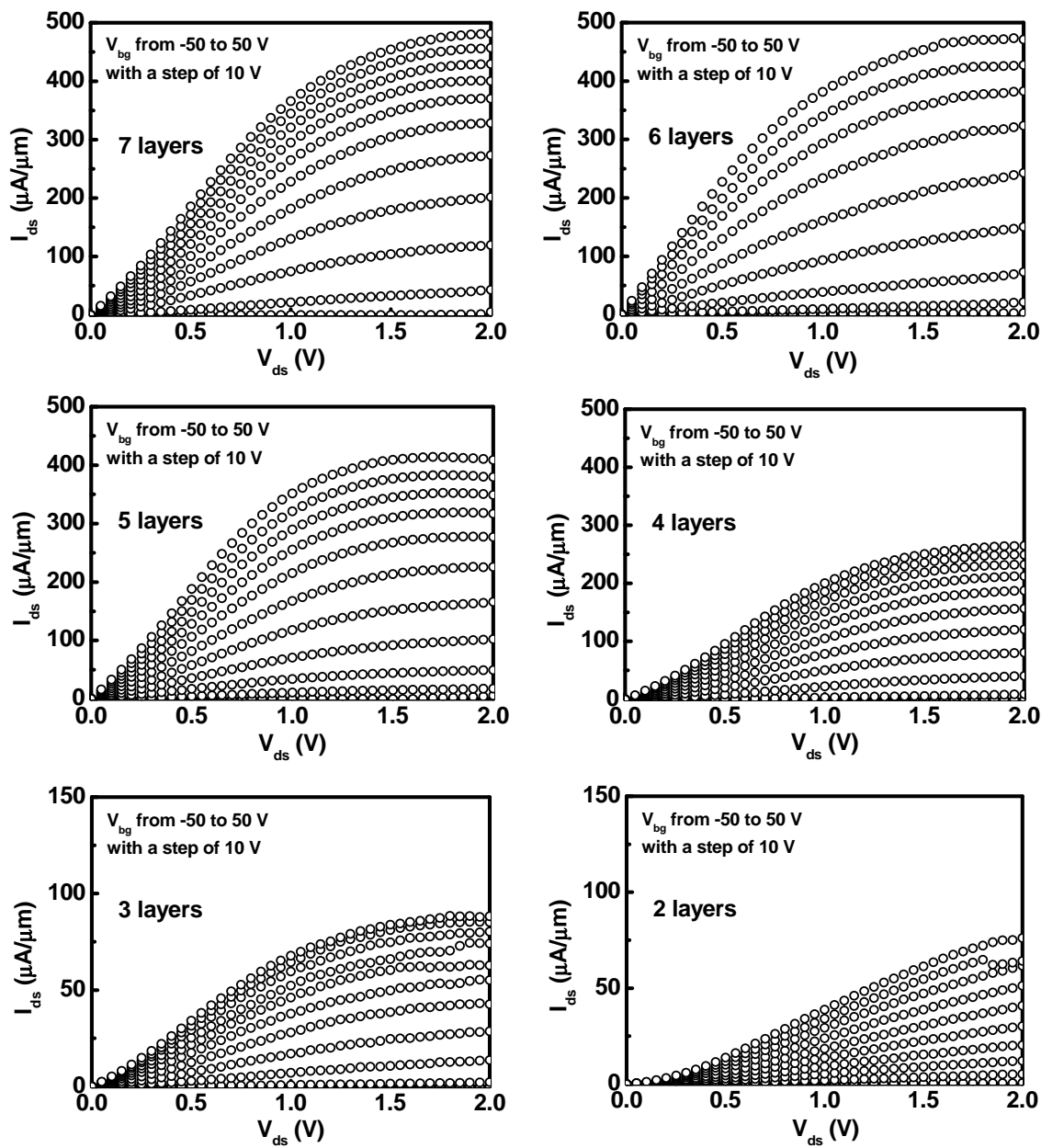

**Figure S2.** $I_d$-$V_d$ of the 100 nm Cl-doped $MoS_2$ FETs with 2-7 layers.

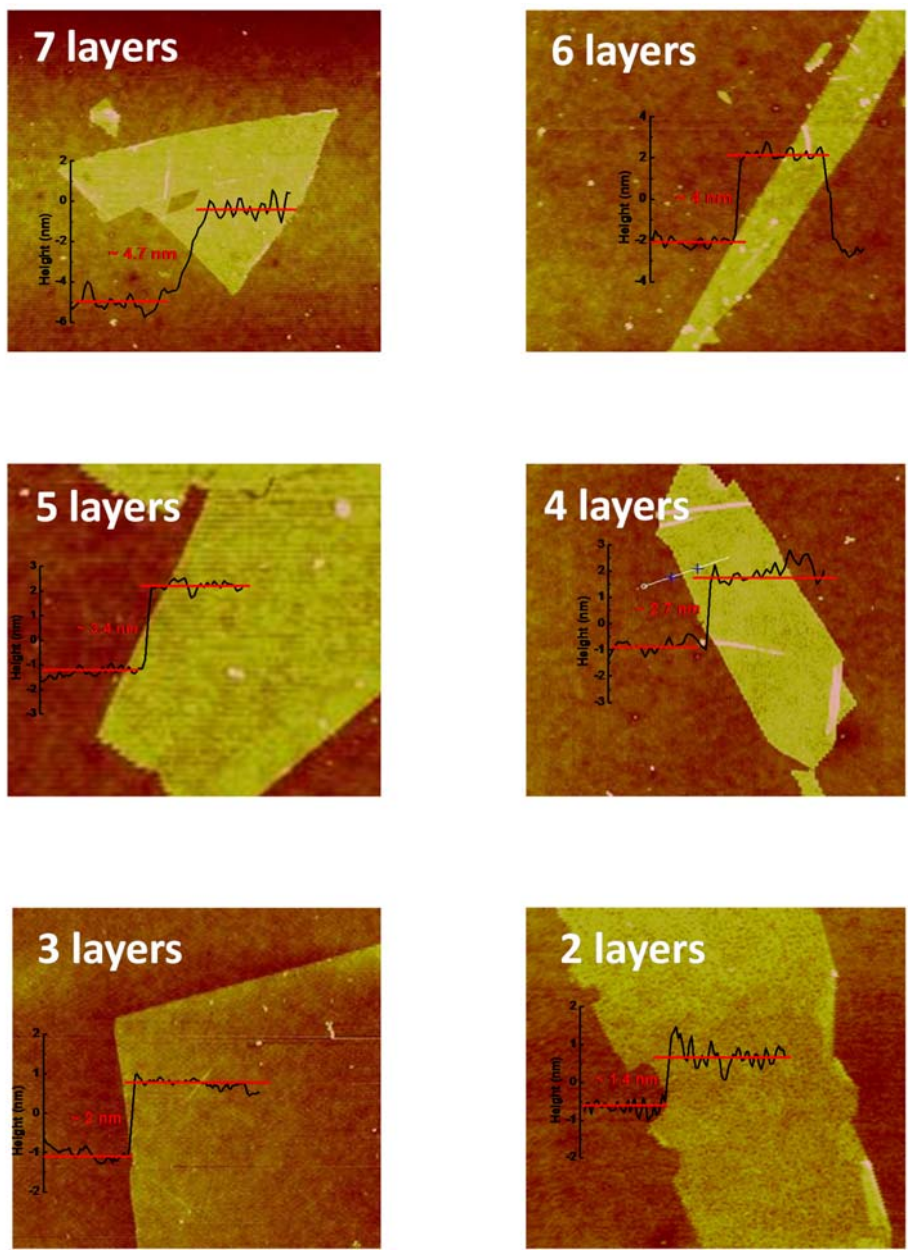

**Figure S3.** AFM images of 2-7 layer MoS$_2$ flakes with the measured different thickness.

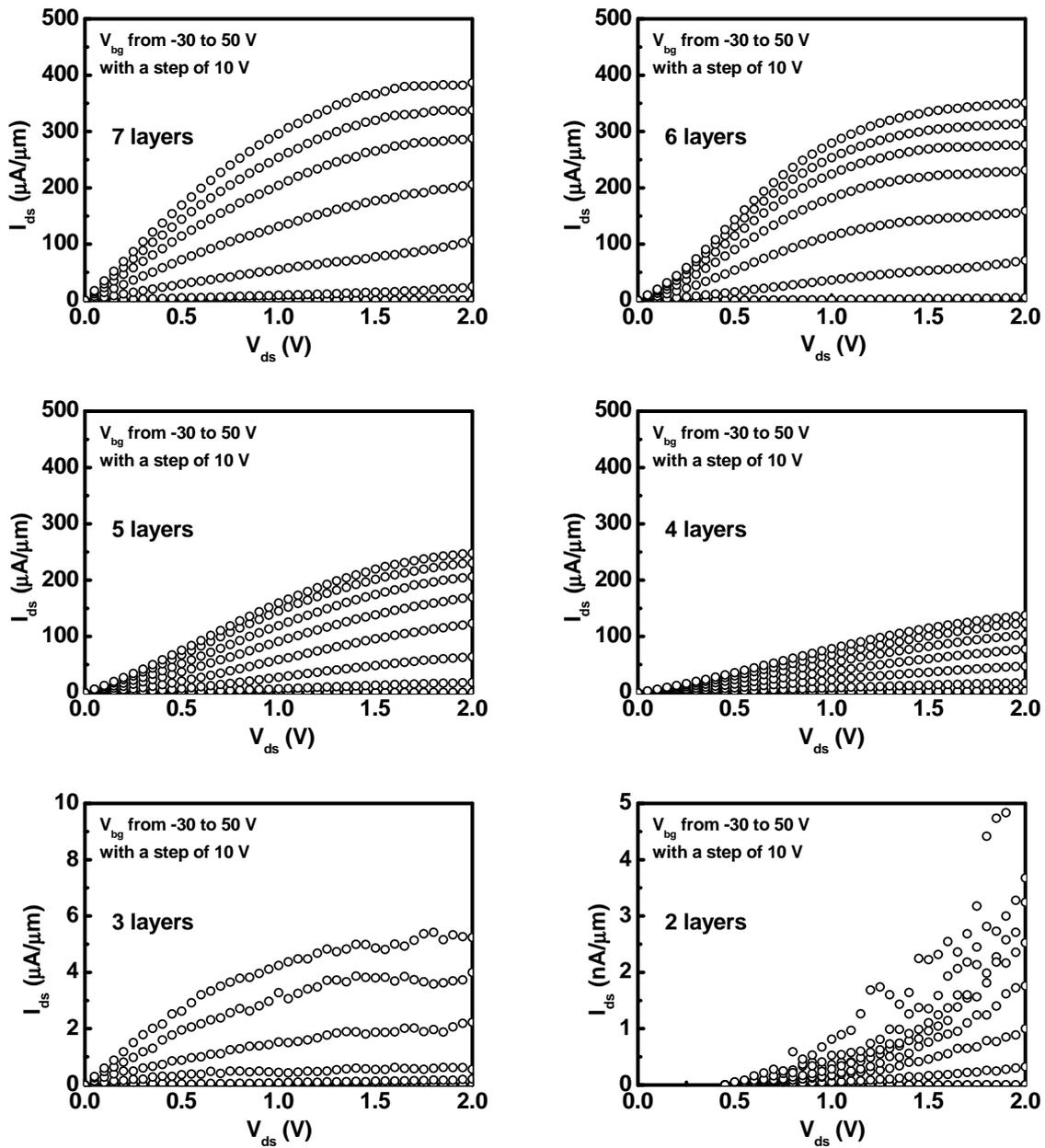

**Figure S4.** $I_d$-$V_d$ of the 100 nm Cl-doped $WS_2$ FETs with 2-7 layers.

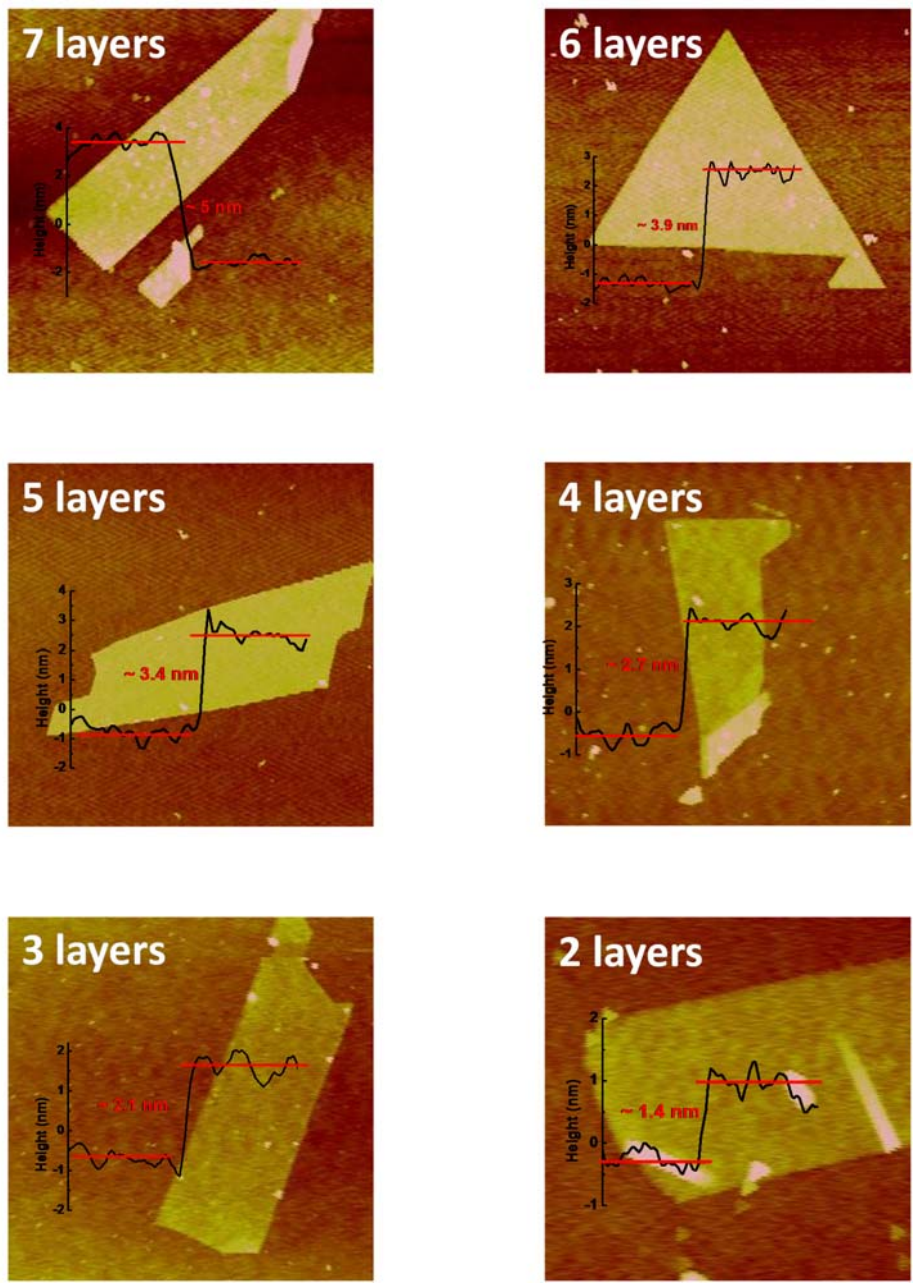

**Figure S5.** AFM images of 2-7 layer $WS_2$ flakes with the measured different thickness.

**Hysteresis data**

Figure S6 shows the $I_d$-$V_g$ hysteresis curves with forward and backward voltage sweeps for devices with different channel lengths (from 100 nm to 1 μm). In order to exclude the effect of $H_2O$ and $O_2$ on hysteresis, $N_2$ environment is used during the measurement. The long channel devices show reasonable hysteresis due to the thick $SiO_2$ and a 10 V threshold voltage ($V_{th}$) shift. For short channel device, the $V_{th}$ shift increases to 20 V. It seems that the hysteresis increases due to the increasing lateral electric field when the channel length is scaled ($E = \frac{V_{ds}}{L_{ch}}$). Except for the $V_{th}$ shift, the devices show similar ON and OFF currents with both forward and backward sweeps.

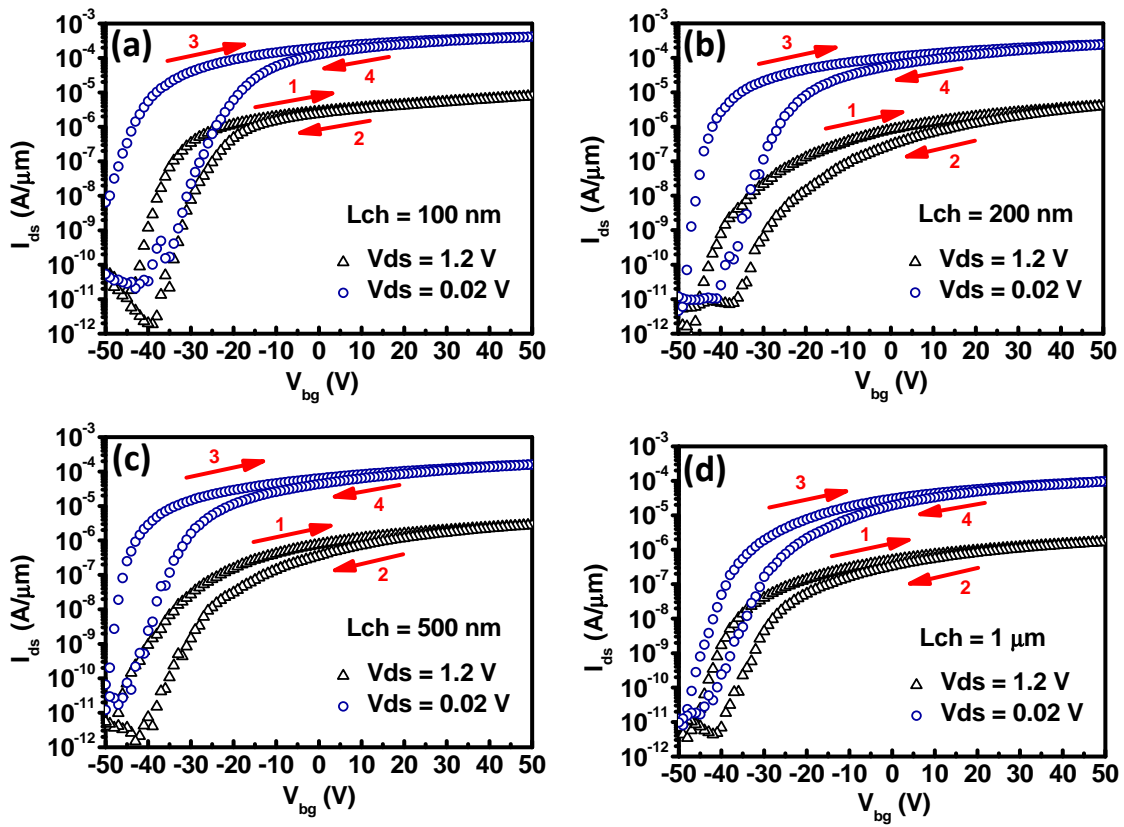

**Figure S6.** Dual sweep $I_d$-$V_g$ curves of the Cl-doped $MoS_2$ FETs with a channel length of (a) 100nm, (b) 200 nm, (c) 500 nm, and (d) 1 μm.

## Stability of Cl-doped FETs in vacuum and air

The stability of Cl-doped FETs both in vacuum and air is studied. The MoS$_2$ device is very stable in vacuum or air for four weeks, as can be seen from Figure S7. Except for a minor $V_{th}$ shift, the transfer characteristics curves of the devices after 4 weeks remain the same as the fresh ones. The off current does not increase either. The linear $I_d$-$V_d$ curves also show that the contact is still low resistive although the drain current decreases a little bit. Figure S8 shows the TLM resistance of the WS$_2$ FETs for both fresh devices and devices exposed in air for four days. The $R_c$ increased from 0.7 to 1.2 kΩ·μm while the channel sheet resistance increases from 11 to 21 kΩ·μm. The degradation of device in air is possibly due to the interaction with O$_2$ and/or H$_2$O. Since the contact region is covered and protected by the contact metal, the degradation of the contact is less severe than that of the uncovered channel region.

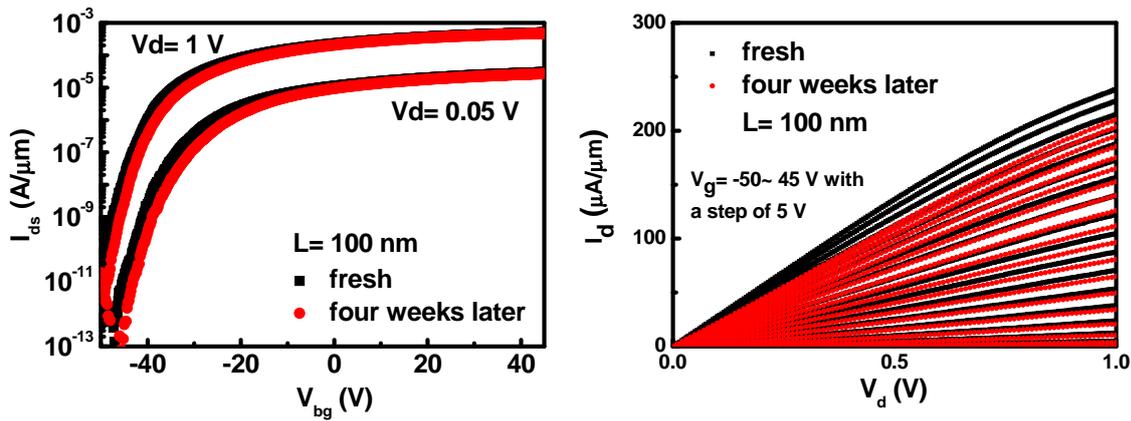

**Figure S7.** Stability of the Cl-doped MoS$_2$ FETs in vacuum after 4 weeks.

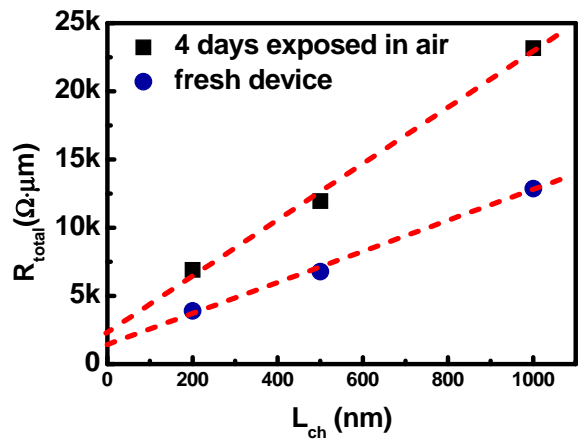

**Figure S8.** Stability of the Cl-doped WS$_2$ FETs in air after 4 days.